\documentclass[a4paper,11pt,twocolumn]{extarticle}
\usepackage{stfloats}
\usepackage{float}

\usepackage{multicol}
\usepackage{amsmath}
\usepackage[T1]{fontenc}
\usepackage[dvipsnames]{xcolor}
\usepackage[section]{placeins}
\usepackage{xcolor}
\usepackage[toc,page,header]{appendix}
\usepackage{graphicx}
\usepackage[font=small,labelfont=bf]{caption}
\usepackage{multirow}
\usepackage{caption}
\usepackage{enumitem}
\usepackage{subcaption}
\usepackage[utf8]{inputenc}
\usepackage{chngcntr}
\usepackage[style = numeric-comp, sorting = none, maxnames=10 ]{biblatex}
\usepackage{authblk}

\bibliography{biblio.bib}
\usepackage[width=210.00mm, height=297.00mm, left=1.7cm, right=1.7cm, top=2cm, bottom=2cm]{geometry}
\usepackage{ amssymb }
\usepackage{verbatim}
\usepackage{caption}
\usepackage{subcaption}

\title{A study on the effect of input data length on deep learning based magnitude classifier}

\date{}

\author[1,4]{Megha Chakraborty}
\author[1]{Wei Li}
\author[1,2]{Johannes Faber}
\author[1,4]{Georg Rümpker}
\author[1,2,3,5]{Horst Stoecker}
\author[1 *]{Nishtha Srivastava}

\affil[1]{Frankfurt Institute for Advanced Studies, 60438 Frankfurt am Main, Germany}
\affil[2]{Institute for Theoretical Physics, Goethe Universität, 60438 Frankfurt am Main, Germany}
\affil[3]{Xidian-FIAS international Joint Research Center, Giersch Science Center,
D-60438 Frankfurt am Main, Germany}
\affil[4]{Institute of Geosciences, Goethe-University Frankfurt, 60438 Frankfurt am Main, Germany}
\affil[5]{GSI Helmholtzzentrum für Schwerionenforschung GmbH, 64291 Darmstadt, Germany}
\affil[*]{srivastava@fias.uni-frankfurt.de}

\renewcommand{\baselinestretch}{1.2} 

\begin{document}
\renewcommand{\baselinestretch}{1.2} 
\maketitle

\renewcommand{\baselinestretch}{1.2} 

\begin{abstract}
The rapid characterisation of earthquake parameters such as its magnitude is at the heart of Earthquake Early Warning (EEW). In traditional EEW methods the robustness in the estimation of earthquake parameters have been observed to increase with the length of input data. Since time is a crucial factor in EEW applications, in this paper we propose a deep learning based magnitude classifier and, further we investigate the effect of using five different durations of seismic waveform data after first P-wave arrival-- 1s, 3s, 10s, 20s and 30s. This is accomplished by testing the performance of the proposed model that combines Convolution and Bidirectional Long-Short Term Memory units to classify waveforms based on their magnitude into three classes-- "noise", "low-magnitude events" and "high-magnitude events". Herein, any earthquake signal with magnitude equal to or above 5.0 is labelled as high-magnitude. We show that the variation in the results produced by changing the length of the data, is no more than the inherent randomness in the trained models, due to their initialisation.
\end{abstract}

\section{Introduction}
The earthquake magnitude, defined as a logarithmic measure of the relative strength of an earthquake, is one of the most fundamental parameters in its characterisation\cite{Mousavi}. The complex nature of the geophysical processes affecting earthquakes makes it very difficult to have a single reliable measure for its size \cite{Kanamori} and hence, magnitude values measured in different scales often differ by more than 1 unit. This is especially true for larger events due to saturation effects \cite{sat1,sat4}. Owing to above-mentioned reasons and the empirical nature of majority of the magnitude scales, it is one of the most difficult parameters to estimate \cite{empirical1, empirical2}.

Some of the classical approaches to obtain first estimates of earthquake magnitude have used empirical relations for parameters such as predominant period $\tau_ p^{max}$ \cite{Nakamura,allen}, effective average period $\tau_c$  \cite{kanamori2005, mag_EEW} in the frequency domain and  parameters such as peak displacement ($P_d$) \cite{wuzhao, mag_EEW} in the amplitude domain calculated from the initial 1-3 seconds of P-waves. These relations form the basis of existing Earthquake Early Warning (EEW) systems in Japan, California, Taiwan etc. (\cite{EEW1} and the references therein). The accuracy of such estimates have been shown to increase with the duration of data used to calculate them \cite{ziv}. 

The recent developments in the area of deep learning \cite{DL}, combined with the availability of affordable high-end computational power through GPUs, have led to state-of-the-art results in image recognition \cite{imagenet,imrec}, speech recognition \cite{speech1,speech2} and natural language processing \cite{nlp1,nlp2}. In fields such as seismology, where the volume of available data has increased exponentially over the last decades \cite{kong}, deep learning has achieved great success in tasks such as seismic phase picking \cite{pick,  liao2021arru, epick}, event detection \cite{wang, det4, Meier}, magnitude estimation \cite{Mousavi}, event location characterisation \cite{loc1,loc2,loc3}, and first motion polarity detection \cite{polarity}. 

Considering that timeliness is of the essence in rapid earthquake characterisation, it becomes important to find an optimum duration for the input data, that can provide a reliable and statistically significant estimate for various earthquake parameters while using minimum amount of P-wave data. In this study, we present a deep learning model to perform time-series multiclass classification \cite{tsc, multiclass} that classifies seismic waveforms as -- "noise, "low-magnitude" or "high-magnitude". Here a local magnitude of 5.0 is taken to be the boundary between the low-magnitude and high-magnitude classes. We further investigate the effect of using different lengths of data on the model performance. Please note, that the boundary of 5.0 is arbitrarily chosen, and can be modified depending on the purpose of the model and the local geology (which influences the correlation between earthquake magnitude and intensity). The boundary in itself  does not influence the model performance. Unlike \cite{saad}, which uses data from three seismic station to characterise different earthquake parameters, the model discussed in this paper only uses three-component data from a single station. 
\section{Methodology}
\subsection{Data Used}
We use data from the STanford EArthquake Dataset (STEAD) \cite{stead} to train and test our model. STEAD is a high-quality bench-marked dataset created for machine learning and deep learning applications and contains seismic event and noise waveforms of duration 1 minute recorded by over 2,500 seismic stations across the globe. The waveforms have been detrended and filtered with a bandpass filter between 1.0 to 40.0 Hz, followed by a resampling at 100Hz. A metadata consisting of 35 attributes for earthquake traces and 8 attributes for noise traces is provided by the authors. 

To ensure consistency in magnitude we only use traces for which the magnitude is provided in `ml' scale (as this is the case for most of the traces in the dataset). We also discard traces with signal-to-noise ratio less than 10dB for quality control. We divide the noise and earthquake traces into training, validation and test sets in the ratio 60:10:30. Care is taken to make sure that the three aforementioned datasets are non-overlapping. This means, that traces corresponding to a particular earthquake (represented by the `source\_id' attribute) but recorded at different stations are included in only one of the three sets. For noise traces, recordings from a particular seismic station are included in only one of the three sets. 

In this paper, we propose a classifier model for rapid earthquake characterisation. Furthermore, we investigate the effect of using different lengths of data after the first P-arrival (1s, 3s, 10s, 20s and 30s) on the performance of this classifier model. In each case the P-wave data is preceded by 2.8-3.0 seconds of pre-signal noise, so the model can learn the noise characteristics of the station \cite{team}. The data labels 0, 1, and 2 are used to denote the classes noise, low-magnitude and high-magnitude, respectively.

\begin{figure}[t]
    \centering
    \includegraphics[scale = 0.5]{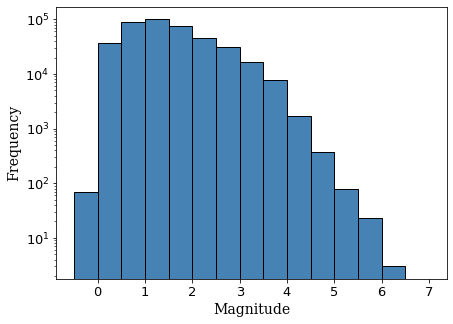}
    \caption{Original distribution of local magnitudes in the chunk of STEAD \cite{stead} data used for training.} 
    \label{mag_dist}
\end{figure}

As mentioned earlier, we take a local magnitude 5.0 to be the decision boundary between high-magnitude and low-magnitude events. However, the training dataset originally has a magnitude distribution as shown in Figure \ref{mag_dist}; this would lead to a high imbalance between the low-magnitude and high-magnitude classes (a ratio of nearly 3300:1). It is widely agreed by the Machine Learning community that most classifiers assume an equal distribution between the different classes.\cite{imbalance} Although examples from some domains where models perform reasonably well even in highly imbalanced datasets, show that there are other factors at play, imbalanced datasets not only are a major hindrance in the development of good classifiers but can also lead to misleading evaluations of the accuracy of the model \cite{imbalance}. To tackle this \textit{imbalance problem} we apply resampling of the data \cite{balance} as follows:
\begin{itemize}
    \item Events with magnitude equal to or above 5.0 are represented 20 times in the dataset, by using a \textit{shifting window} starting from 300 samples to 280 samples before the first P-arrival sample, the window being shifted by 2 samples for each representation. Each of these traces are also \textit{flipped}, i.e. their polarity is reversed, since it does not affect the magnitude information of the data. Such data augmentation techniques used for images have also been found to be useful for time series data \cite{imbalance, aug}.
    \item For low-magnitude events the following strategy of random-undersampling is adopted:
    \begin{enumerate}
         
    \item All events with magnitude between 4.5 and 5.0 are used.
    \item 1/3$^\text{rd}$ of events with magnitude between 4.0 and 4.5 are used.
    \item 1/50$^\text{th}$ of events with magnitude between 2.0 and 4.5 are used.
    \item 1/100$^\text{th}$ of events with magnitude less than 2.0 are used.
    \end{enumerate}
    \item 1/25$^\text{th}$ of the available noise traces are used.
\end{itemize}

\begin{figure}[t]
    \centering
    \includegraphics[scale = 0.6]{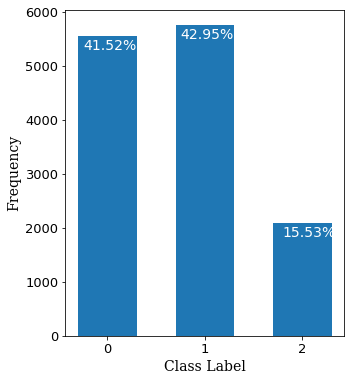}
    \caption{The distribution of classes in the training dataset obtained by under-sampling noise and low-magnitude data and applying data-augmentation to high-magnitude events. Classes 0,1 and 2 represent \textit{`noise'}, \textit{`low-magnitude'} and \textit{`high-magnitude'} data, respectively. A similar distribution of classes is seen in the validation and test datasets as well.}
    \label{class}
\end{figure}
\begin{figure*}[!h]
    \centering
    \includegraphics[scale = 0.8]{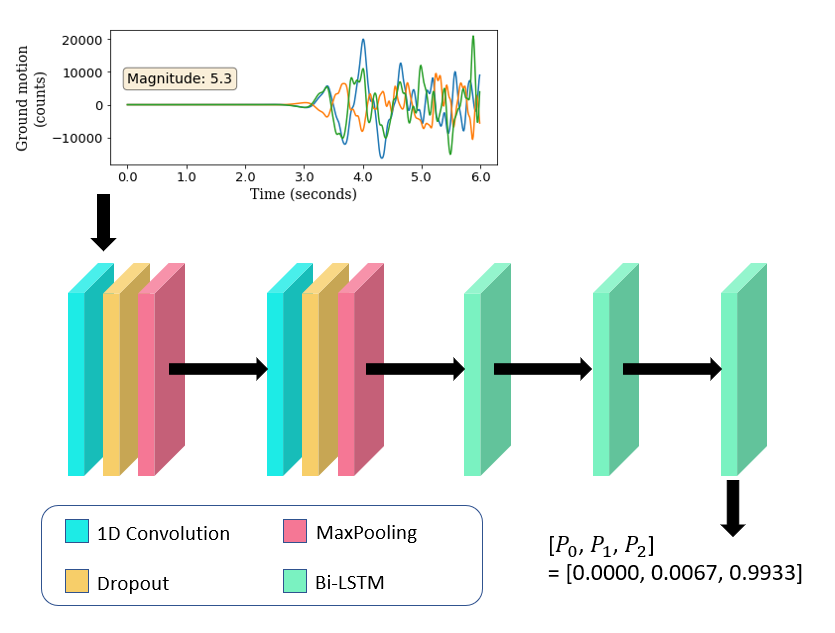}
    \caption{The architecture of the model used to perform the 3-class classification. The input to the model is 3-component seismic waveform data from a single station. The example shown here corresponds to the case where 3 seconds of P-wave data is used (the total length of data is, thus, 6 seconds). The 1D Convolution layers have a kernel size of 4 and 8 filters each; the drop rate for each Dropout layer is 0.2, and each MaxPooling layer reduces the size of the data by a factor of 4; the Bi-LSTM layers have dimensions of 256, 256 and 128, respectively. The final layer is a Softmax layer, that outputs the probability of the trace belonging to classes 0 (noise), 1 (low-magnitude) and 2 (high-magnitude), represented here as $P_0$, $P_1$ and $P_2$ respectively. In this case a probability of 0.9933 is assigned to class 2, for an event with magnitude 5.3; thus, this is a case of correct classification.} 
    \label{arch}
\end{figure*}
Note that special care is taken to include more events close to the decision boundary, so that the model can learn to differentiate between events of magnitude say, 4.0 to 5.0 which is more difficult compared to differentiating between events of magnitude say, 2.0 and 5.0. The corresponding distribution of the different classes is shown in Figure \ref{class}. The validation and test datasets follow a similar distribution. As one can see, in spite of the resampling techniques employed, the high-magnitude class is still under-respresented in the dataset, as compared to the other two classes. So we apply a class-weight \cite{balance} of 1:1:10 (chosen, experimentally) for classes 0,1 and 2 while training the model. The data is used without instrument response removal. Unlike \cite{cnq} we do not normalise the data. Only the waveform information is provided to the model.

\subsection{Model Architecture and Model Training}
The model architecture \cite{egu} consists of two sets of 1D Convolution \cite{conv1d}, Dropout \cite{drop} and MaxPooling\cite{maxpool}, followed by three bidirectional Long-Short Term Memory (LSTM) \cite{lstm} layers; the final layer is a Softmax layer \cite{softmax} which gives a three-element array of the form $[P_0, P_1, P_2]$, where $P_i$ is the probability of the waveform belonging to the class $i$ (Figure \ref{arch}).

The model is trained using Adam optimiser \cite{adam}, Categorical Crossentropy \cite{crossentropy} loss and a batch size of 256. Early stopping \cite{earlystopping} is used to prevent overfitting, whereby the validation loss is monitored and the training stops when there is no reduction in it for 20 consecutive epochs. We start with a learning rate of $10^{-3}$ and reduce it by a factor of 10 if the validation loss does not reduce for 15 consecutive epochs until it reaches $10^{-6}$. The model for the epoch corresponding to the lowest validation loss is retained. 

\begin{figure*}[tb]
    \centering
    \includegraphics[scale = 0.55]{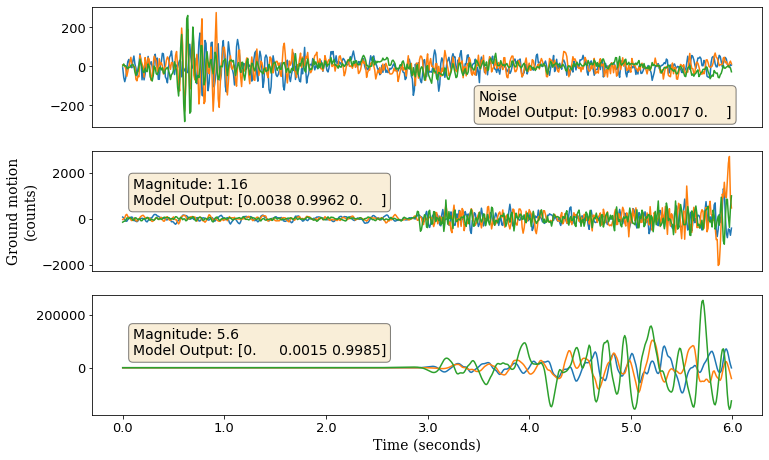}
    \caption{Examples of waveforms that have been correctly classified. In each case the highest probability corresponds to the respective class.} 
    \label{correct}
\end{figure*}

\section{Results}
To analyse the effect of different lengths of data on the performance of the classifier model, we use the metrics listed below to evaluate the model performance. The metrics are calculated in terms of true positives (TP), true negatives (TN), false positives (FP) and false negatives (FN).

\begin{itemize}
    \item \textbf{Accuracy}: The accuracy of a classifier is the proportion of testing samples that are correctly classified. Mathematically, it can be defined as follows:
    \begin{equation}
        \text{Accuracy} = \frac{TP + TN}{TP + FP + TN + FN}
    \end{equation}
    \item \textbf{Precision}: This is the ratio of the number of times the model \textit{correctly} predicts a class to the total number of times it predicts that class. Mathematically it is defined as:
    \begin{equation}
        \text{Precision} = \frac{TP}{TP + FP}
    \end{equation}
    \item \textbf{Recall}: This is the ratio of the number of times the model correctly predicts a class to the total number occurences of that class in the dataset. Mathematically it is defined as:
    \begin{equation}
        \text{Recall} = \frac{TP}{TP + FN}
    \end{equation}
\end{itemize}

\begin{figure*}[t]
    \centering
    \includegraphics[scale = 0.6]{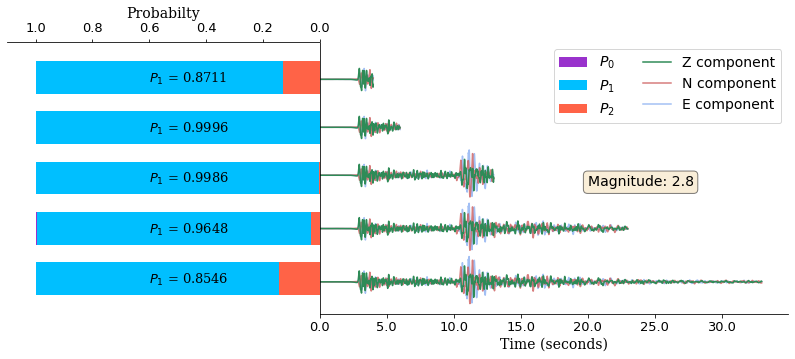}
    \caption{Softmax probabilities for different input lengths of the same , predicted by the models trained on the corresponding lengths of data. The waveform used here corresponds to an event of magnitude 2.8, although the maximum probability corresponds to class 1, the values of these probabilities are different for different data lengths, and there is no clear dependence between the length of the data and this probability.} 
    \label{lengths}
\end{figure*}

\begin{figure*}[!htp]
\begin{subfigure}{.5\textwidth}
      \centering
      \includegraphics[width=\linewidth]{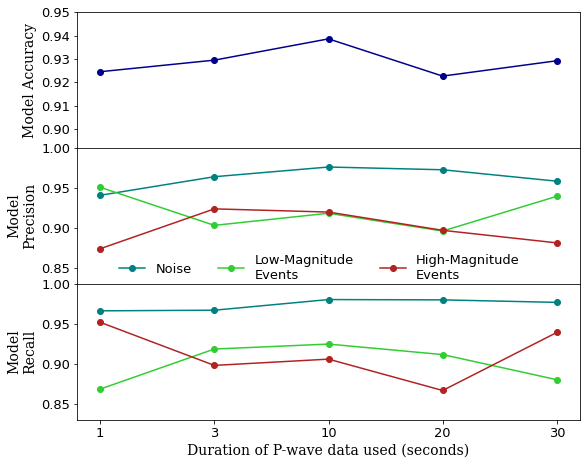}  
        \caption{}
        \label{duration}
    \end{subfigure}
    \begin{subfigure}{.5\textwidth}
      \centering
      \includegraphics[width=\linewidth]{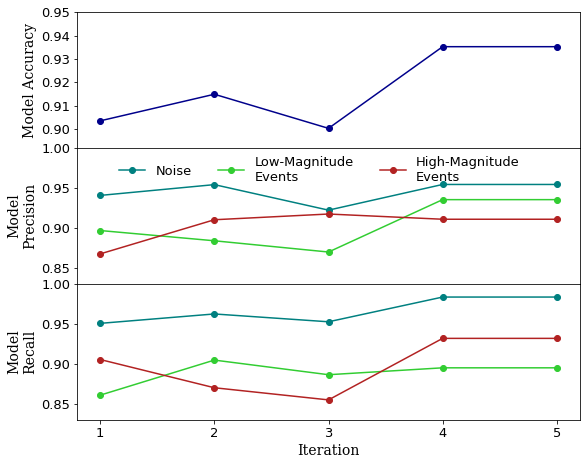}  
        \caption{}
         \label{iteration}
    \end{subfigure}
    \caption{(a) Variation in classifier model performance when different duration of P-wave data are used; (b) Variation in the classifier model performance when the same model is re-trained on the same data (in this case 3 seconds of P-wave data used) five times. This shows that the variation in the two cases are comparable.}
    \label{fig}
\end{figure*}

Figure \ref{correct} shows three waveforms, (one from each class) that has been correctly classified. The softmax probalities, as described in section 2.2, are also shown. In each case the highest probability is predicted for the corresponding class. 

Figure \ref{lengths} shows the softmax probalities,  predicted by the model for different lengths of the same waveform. Although the waveform is correctly classified in each case, the predicted probabilities are different and show no dependence on the length of input data. Figure \ref{duration} shows the variation in the model performance with the duration of P-wave data used. We also look at the randomness in the performance when the model is trained on the same data five times (Figure \ref{iteration}), as we do not tune a random seed during model training \cite{rando1, rando2}. Thus, we can see that the variation in the results caused by changing the length of data is comparable to the randomness in the results due to random-initialisation upon re-training the model on the same data.

\begin{figure*}[t]
    \begin{subfigure}{.5\textwidth}
      \centering
      \includegraphics[width=\linewidth]{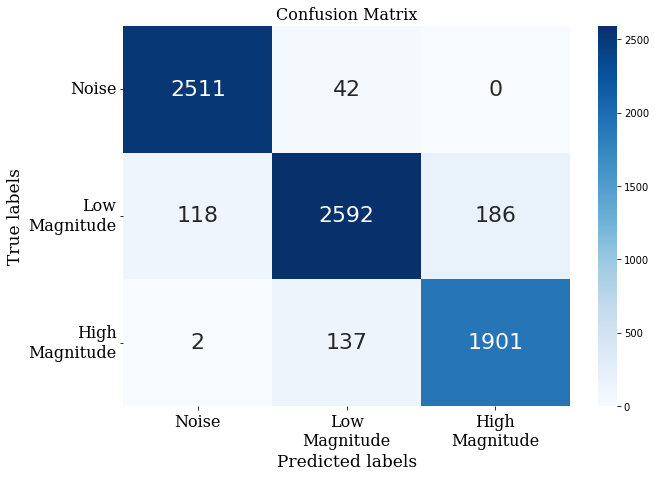} 
        \caption{}
        \label{5a}
    \end{subfigure}
    \begin{subfigure}{.5\textwidth}
      \centering
      \includegraphics[width=\linewidth]{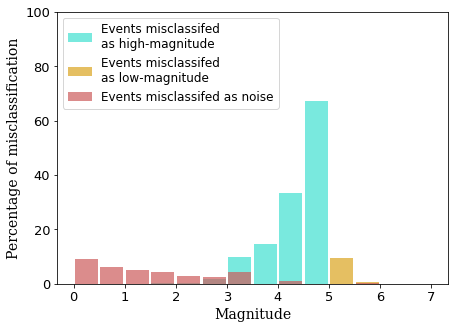} 
        \caption{}
         \label{5b}
    \end{subfigure}
    \caption{The classification results for a model trained on the 3 second data. (a) The confusion matrix \cite{Ting2017} for a model trained and tested on the 3 second data. (b) The mis-classification statistics for the same model, for different magnitude values. Note how the highest degree of mis-classification happens close to the decision boundary; the percentage of low-magnitude events classified as high-magnitude is much higher than the percentage of high-magnitude events classified as low-magnitude; this is a result of the class-weights we used while training the model.}
    \label{classifier}
\end{figure*}

Figure \ref{classifier} shows the classification statistics for one of the iterations of the model trained on the 3 second data. One can see that the events classified as noise tend to be of low magnitude, while the mis-classification of low-magnitude events as high-magnitude and vice-versa, is most pronounced at the decision boundary of 5.0. Another important observation is that the degree of misclassification of low-magnitude events is much higher than the reverse case; approximately 65\% of events with magnitude between 4.5 and 5.0 and 35\% of events with magnitude between 4.0 and 4.5 get classified as high magnitude, while less than 10\% of events with magnitude between 5.0 and 5.5 are classified as low-magnitude; this is intentional as a missed alarm is considered more dangerous than a false alarm in this context \cite{EEW2} and is achieved by giving the high-magnitude class more weight during model training.

\section{Conclusion}
In this study, we present a deep learning model
that classifies seismic waveform into three-classes:
\textit{noise}, \textit{low-magnitude} events and \textit{high-magnitude}
events, with events having local magnitude equal
to or above 5.0 categorised as `high-magnitude’.
We investigate the effect of using different duration of P-wave data to perform the said task and
demonstrate that changing the length of the waveform has no significant effect on the model performance. We also find that the model classifies
most the data above a magnitude of 4.5 as high-magnitude, even though the decision boundary is
chosen at 5.0, due to the higher class weight assigned to high-magnitude events. We obtain an
overall accuracy ranging between 90.04\% and 93.86\% (which is comparable to the magnitude classification accuracy of 93.67\% achieved by \cite{saad} using data from three seismic stations).

\section{Acknowledgement}
This research is supported by the “KINachwuchswissenschaftlerinnen" - grant SAI
01IS20059 by the Bundesministerium für Bildung
und Forschung - BMBF. Calculations were performed at the Frankfurt Institute for Advanced
Studies’ new GPU cluster, funded by BMBF for
the project Seismologie und Artifizielle Intelligenz
(SAI).
 
\printbibliography

\end{document}